\documentclass[10p,twocolumn]{aastex61}

\begin{document}

\title{Wavelength does not equal pressure: vertical contribution functions and their implications for mapping hot Jupiters}

\author{Ian Dobbs-Dixon}
\affil{Department of Physics, NYU Abu Dhabi PO Box 129188 Abu Dhabi, UAE}
\author{Nicolas B. Cowan}
\affil{Department of Physics and Department of Earth \& Planetary Sciences, McGill University, 3550 rue University, Montr\'eal, QC H3A 2A7, Canada}

\begin{abstract}
Multi-band phase variations in principle allow us to infer the
longitudinal temperature distributions of planets as a function of
height in their atmospheres. For example, 3.6~$\mu$m emission
originates from deeper layers of the atmosphere than 4.5~$\mu$m due to
greater water vapor absorption at the longer wavelength. Since heat
transport efficiency increases with pressure, we expect thermal phase
curves at 3.6~$\mu$m to exhibit smaller amplitudes and greater phase
offsets than at 4.5~$\mu$m ---this trend is not observed.  Of the
seven hot Jupiters with full-orbit phase curves at 3.6 and 4.5~$\mu$m,
all have greater phase amplitude at 3.6~$\mu$m than at 4.5~$\mu$m,
while four of seven exhibit a greater phase offset at 3.6~$\mu$m.  We
use a 3D radiative-hydrodynamic model to calculate theoretical phase
curves of HD~189733b, assuming thermo-chemical equilibrium. The model
exhibits temperature, pressure, and wavelength dependent opacity,
primarily driven by carbon chemistry: CO is energetically favored on
the dayside, while CH$_4$ is favored on the cooler nightside.
Infrared opacity therefore changes by orders of magnitude between day
and night, producing dramatic vertical shifts in the
wavelength-specific photospheres, which would complicate eclipse or
phase mapping with spectral data.  The model predicts greater relative
phase amplitude and greater phase offset at 3.6~$\mu$m than at
4.5~$\mu$m, in agreement with the data. Our model qualitatively
explains the observed phase curves, but is in tension with current
thermo-chemical kinetics models that predict zonally uniform
atmospheric composition due to transport of CO from the hot regions of
the atmosphere.
\end{abstract}

\section{Introduction}
If a planet has a sufficiently large day-night temperature contrast,
then it will exhibit thermal phase variations: it will appear brighter
in the infrared when we see its dayside than its nightside.
In practice, this condition holds for short-period planets because
tidal forces tend to make them rotate synchronously, with one
hemisphere always facing the host star, and the other perpetually in
the dark \citep[e.g.,][]{Dobbs-Dixon2004}.  
Regardless of their underlying rotational state, thermal phase measurements indicate that short period planets have day--night temperature contrasts of hundreds
to thousands of Kelvin \citep{cowan2011statistics, perez2013atmospheric,
  schwartz2015balancing, komacek2016atmospheric, Schwartz_2017}.

As a consequence of day-to-night heat transport, the
hottest point on the planet may be displaced from the permanent
sub-stellar location. 
This symmetry breaking occurs
because even a tidally locked planet rotates in an inertial frame. The resulting coriolis forces couple to the day-night temperature differential to accelerate a super-rotating circumplanetary jet \citep{showman2011,
  tsai2014}. Atmospheric circulation models of tidally locked hot
Jupiters almost uniformly predict the atmosphere to be dominated by a
broad super-rotating equatorial jet \citep[e.g.,][and references
  therein]{heng2014atmospheric}. This leads to the general prediction
that their thermal phase variations will peak prior to superior
conjunction, when regions east of the substellar longitude are facing
the observer (east is defined to be in the direction of the planet's
rotation).  This phase offset has indeed been observed for many hot
Jupiters, starting with HD~189733b
\citep{knutson2007map,knutson2009multiwavelength,knutson20123}.

Given the successes of eclipse and phase mapping using \emph{Spitzer},
\emph{Kepler}, and \emph{Hubble}, it is now expected that the James
Webb Space Telescope will enable 3D mapping of the daysides of hot
Jupiters and 2D longitude-pressure maps of their nightsides \citep[for
  a review of exoplanet mapping, see][]{Cowan_Fujii_2017}.
Different wavelengths have different opacities and hence probe
different pressures in the atmosphere: the optical depth to the
top-of-atmosphere is proportional to the mass of overlying gas, as
is the pressure, so optical depth and pressure should be intimately
linked, \emph{provided that the opacity spectrum is roughly constant
  throughout the entire atmosphere}.

In Section~\ref{sec:data} we discuss current trends in \emph{Spitzer}
phase curves of hot Jupiters, including two trends that seem to defy
expectations. In Section~\ref{sec:theory} we use a
radiative-hydrodynamic model to explore the longitudinal dependence of
vertical contribution functions for HD~189733b and find that we can
qualitatively explain the trends in \emph{Spitzer} phase curves. We
discuss our findings and their implications in
Section~\ref{sec:discussion}

\section{Trends in the \emph{Spitzer} Phase Measurements of Hot Jupiters}\label{sec:data}

Thermal phase measurements have been made with the Spitzer Space
Telescope \citep{Werner_2004} for a dozen planets. For consistency, we
only consider full-orbit, continuous phase measurements of planets on
circular orbits acquired with both the 3.6 and 4.5~$\mu$m channels of
the Infrared Array Camera \citep[IRAC;][]{Fazio_2004}.  Our sample
therefore consists of seven planets: WASP-12b
\citep{cowan2012thermalW}, HD~189733b \citep{knutson20123}, WASP-18b
\citep{maxted2013spitzer}, WASP-14b \citep{wong2015W14b}, HAT-P-7b and
WASP-19b \citep{wong20163}, and WASP-43b \citep{stevenson2017spitzer}.

\begin{figure}
  \centering
  \includegraphics[width=1.1\linewidth]{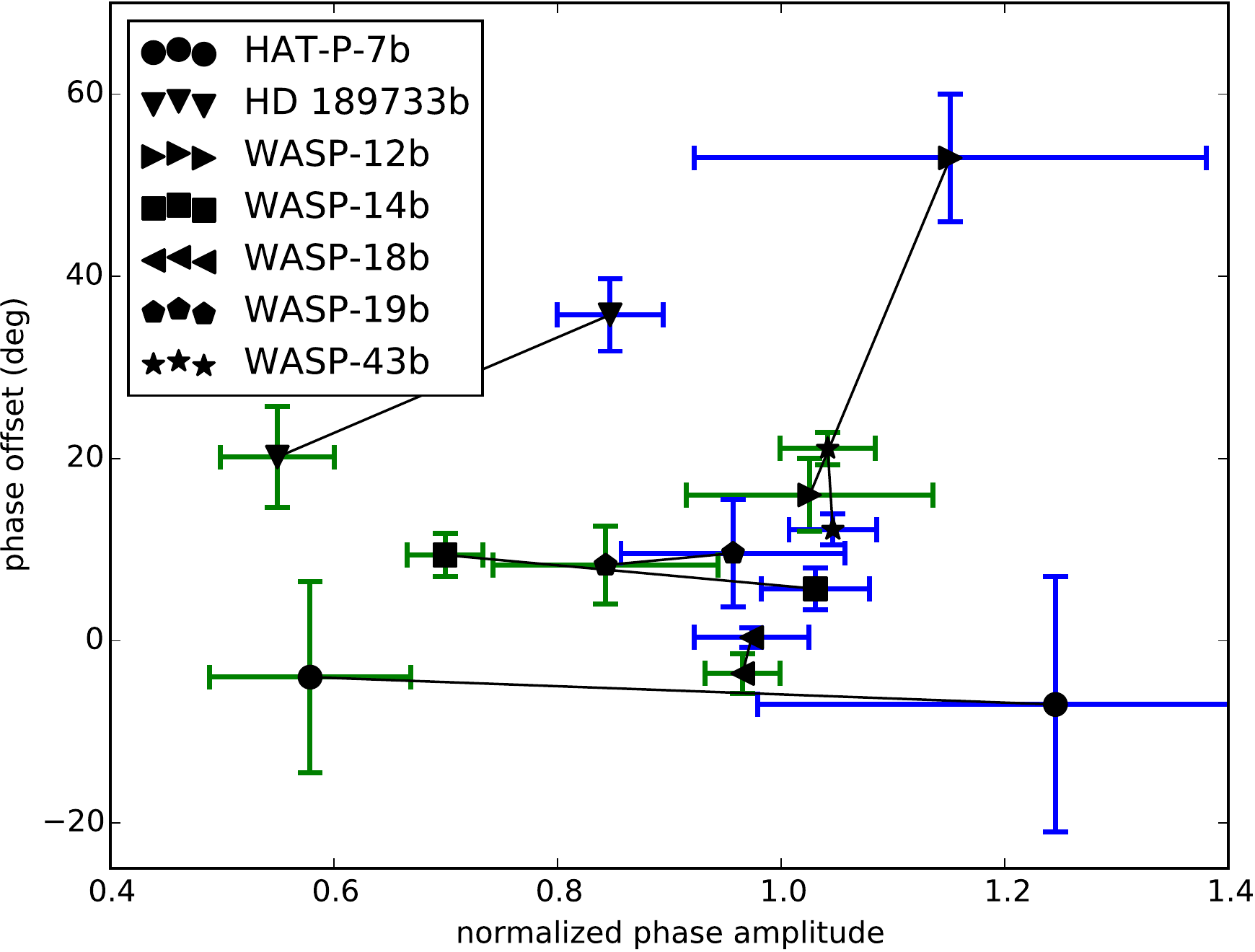}
  \caption{Phase offset versus normalized phase amplitude for six hot
    Jupiters. The normalized phase amplitude is given by $(F_{\rm
      max}-F_{\rm min})/F_{\rm day}$. Blue denotes 3.6~$\mu$m, while
    green is 4.5~$\mu$m; each planet is denoted by different symbols
    and the two wavebands for a given planet are connected by a line
    to guide the eye. The most robust trend is that the normalized
    phase amplitude is always greater at 3.6~$\mu$m than at 4.5~$\mu$m
    (blue points are always to the right of green points).  In
    addition, HD~189733b, WASP-12b, WASP-18b, and WASP-19b exhibit
    greater phase offsets at the wavelength with the greatest
    normalized amplitude. Neither of these trends is expected. Naively
    we would expect the blue points to be found to the upper left of
    the green points.}
  \label{fig:data}
\end{figure}

In Figure~\ref{fig:data} we plot the phase offset (the angular
separation between the peak of the phase curve and superior
conjunction) vs.\ the normalized phase amplitude (the peak-to-trough
amplitude of phase variations divided by the eclipse depth at that
wavelength). To first order, the phase offset is equal to the
longitudinal distance between the sub-stellar point and the zonally
offset hotspot, but they are not interchangeable
\citep[][]{cowan2008inverting,Schwartz_2017}, while the normalized
amplitude is related to the day--night brightness contrast (see
further discussion in Section~\ref{sec:theory}). These data show two
trends: (1) the normalized amplitude of phase variations is
\emph{always} greater at 3.6~$\mu$m than at 4.5~$\mu$m, and (2) the
phase offset is \emph{usually} greater at 3.6~$\mu$m than at
4.5~$\mu$m. Note that the second trend is not nearly as statistically
significant as the first.

One-dimensional clear-sky thermo-chemical equilibrium models of
HD~189733b predict that 3.6~$\mu$m photons originate from deeper in
the atmosphere than 4.5~$\mu$m photons on both the dayside and
nightside \citep[Figure~8 of][]{knutson2009multiwavelength}.  Since
radiative timescales increase with depth, we would expect smaller
amplitude phase variations at 3.6~$\mu$m than at 4.5~$\mu$m---assuming
similar advective timescales (wind speeds) at all pressures.  This
na\"ive prediction is clearly ruled out by the data shown in
Figure~\ref{fig:data}.

Moreover, intuition of damped driven oscillators \citep[and energy
  balance models:][]{cowan2011model,zhang2017effects} suggests that
phase offset and normalized amplitude should be anti-correlated, with
large amplitude phase variations necessarily having a small phase
offset \citep[e.g., Appendix A of][]{Schwartz_2017}. This trend is
also not seen: four of the seven planets (HD~189733b, WASP-12b,
WASP-18b, and WASP-19b) instead exhibit greater phase offsets at the
wavelength with the greatest normalized amplitude. Though the
equilibrium temperatures and surface gravities of these planets vary,
simulations \citep[see][]{heng2014atmospheric} of a wide array of
planets exhibit a dynamical structure comparable to HD~189733b, so we
expect similar behavior.



\section{Predictions from a Radiative-Hydrodynamic Model of HD~189733\MakeLowercase{b}}\label{sec:theory}
To explore these phenomena, we use a radiative hydrodynamic code that
solves the fully compressible Navier-Stokes equations coupled with
wavelength-dependent radiative transfer to simulate the planetary
atmosphere of HD~189733b \citep{Dobbs-Dixon_2013}. The equations are
solved in spherical coordinates with resolution $\{N_r, N_{\phi},
N_{\theta}\}=\{100, 160, 64\}$, where $r$ is the radial distance,
$\phi$ is the longitude, and $\theta$ is the latitude. Transfer of
energy via radiation employs a frequency-dependent two-stream
approximation \citep{Mihalas1978}. The full planetary spectrum is
divided in 30 bins utilizing averaged frequency-dependent opacities
from \cite{Sharp_Burrows_2007}. A broad super-rotating equatorial jet
and counter rotating mid latitude jets are the dominant dynamical
features, similar to many others in the literature. Observable
quantities are calculated by tracing rays through the simulated
atmosphere at 5000 wavelengths between 0.3 and
30~$\mu$m. Qualitatively, the model is seen to match transit, eclipse,
and phase curve observations, though quantitative differences remain
(see below). Further, though we present simulations only of
HD~189733b, as discussed above, the underlying dynamical features
exhibited in this atmosphere are expected to be applicable to a wide
range of planets. More on both the radiative-hydrodynamic simulation
and the method of calculating observable quantities can be found in
\cite{Dobbs-Dixon_2013}.

\begin{figure}
  \centering
  \includegraphics[width=1.0\linewidth]{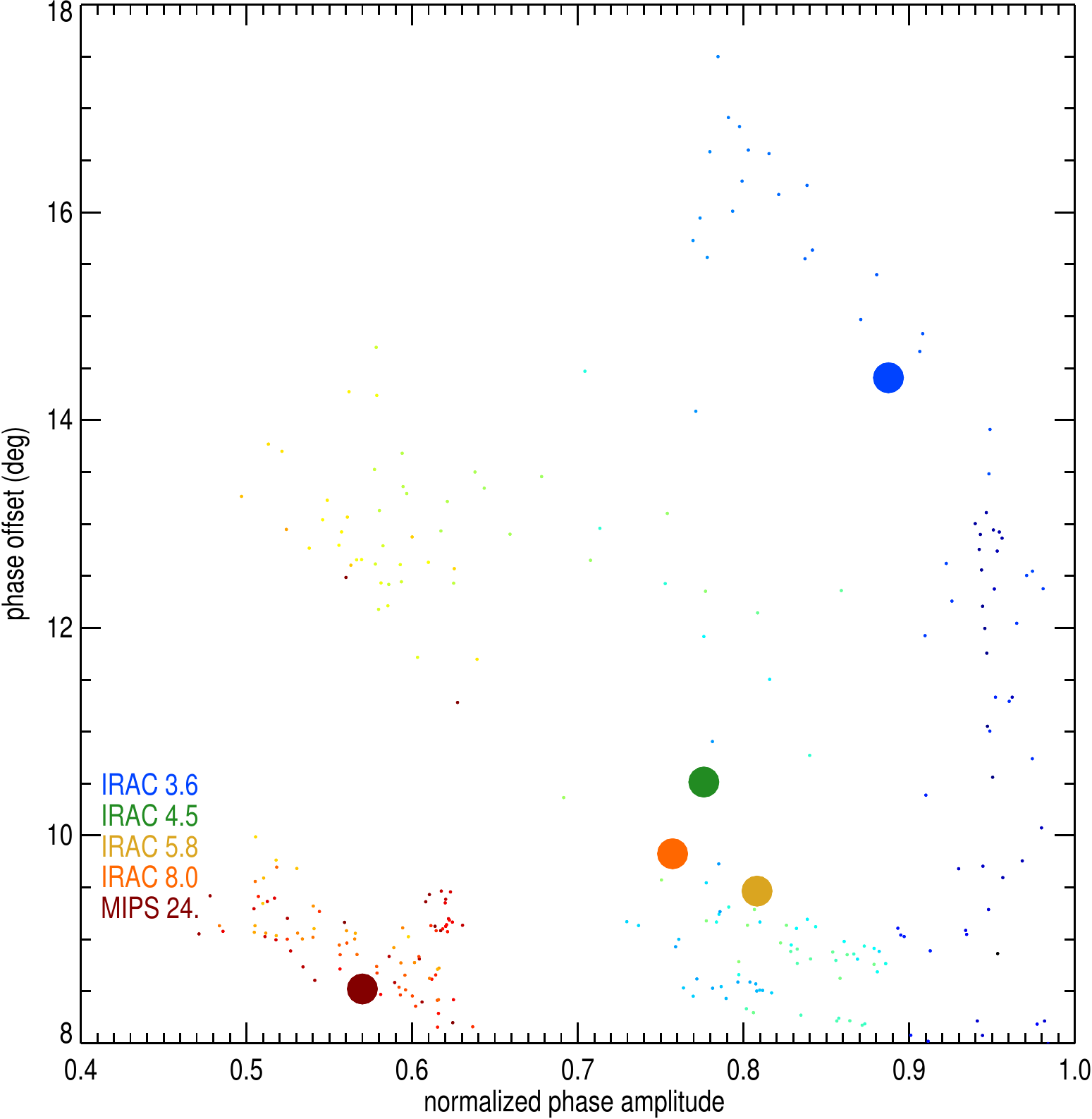}
  \caption{Phase offset vs.\ normalized phase amplitude from the
    radiative-hydrodynamical model of HD~189733b. The normalized phase
    amplitude is given by $(F_{\rm max}-F_{\rm min})/F_{\rm
      day}$. Small dots illustrate values at a subset of the
    individual wavelengths (colored from red to blue with decreasing
    wavelength), while the larger dots are the IRAC and MIPS
    band-averaged quantities. Both quantities are extracted from
    wavelength dependent theoretical phase curves generated by ray
    tracing through the radiative-hydrodynamical model. As the phase
    curves are not sinusoidal, we utilize the simulated fluxes at
    phases near the maximum and minimum to fit for $F_{max}$, the
    offset, and $F_{\rm min}$. $F_{\rm day}$ is the planetary flux
    during eclipse. Consistent with the observations presented in
    Figure~\ref{fig:data}, both the offset and normalized amplitude at
    3.6~$\mu$m are greater than at 4.5~$\mu$m. Offsets at 5.8~$\mu$m,
    8.0~$\mu$m, and 24.0~$\mu$m are smaller and quite similar as they
    are probing the upper regions of the atmosphere (see text). The
    varying phase amplitude for these bands comes from the convoluted
    structure of their contribution functions with longitude, shown in
    Figure~\ref{fig:cf}.}
  \label{fig:theory}
\end{figure}

To compare our simulation to Figure~\ref{fig:data} we calculate both
the phase offset and the normalized phase amplitude. This is done by
first calculating a theoretical phase curve from the model;
essentially a compilation of emission spectra as the sub-observer
longitude moves around the planet as it would throughout an
orbit. Once we have a wavelength dependent phase curve, we bin it
using the IRAC and MIPS band-passes to assemble the phase curves for
any bandpass. From these band averaged phase curves, it is trivial to
fit for maximum flux, minimum flux, and the offset of the maximum
flux. The results are plotted in Figure~\ref{fig:theory}.  Small
colored dots represent a subset of individual wavelengths, while the
band-integrated quantities are denoted by large colored dots.  The
predicted 3.6~$\mu$m phase curve has a greater amplitude and phase
offset than the 4.5~$\mu$m phase curve, as observed for HD~189733b and
many other hot Jupiters (cf.\ Figures~\ref{fig:data} and
\ref{fig:theory}). However, though we match the relative locations of
the 3.6 and 4.5~$\mu$m bands, they do not match quantitatively. This
was recognized in \cite{Dobbs-Dixon_2013}, who noted that the phase
offsets in the model were uniformly under-predicted, suggesting the
need for either a stronger jet or a longer radiative timescale. In
fact, no current 3D models quantitatively predict the phase offset,
with most other models over-predicting it
\citep[e.g.][]{showman2009atmospheric}.

\begin{figure}
  \centering
  \includegraphics[width=1.0\linewidth]{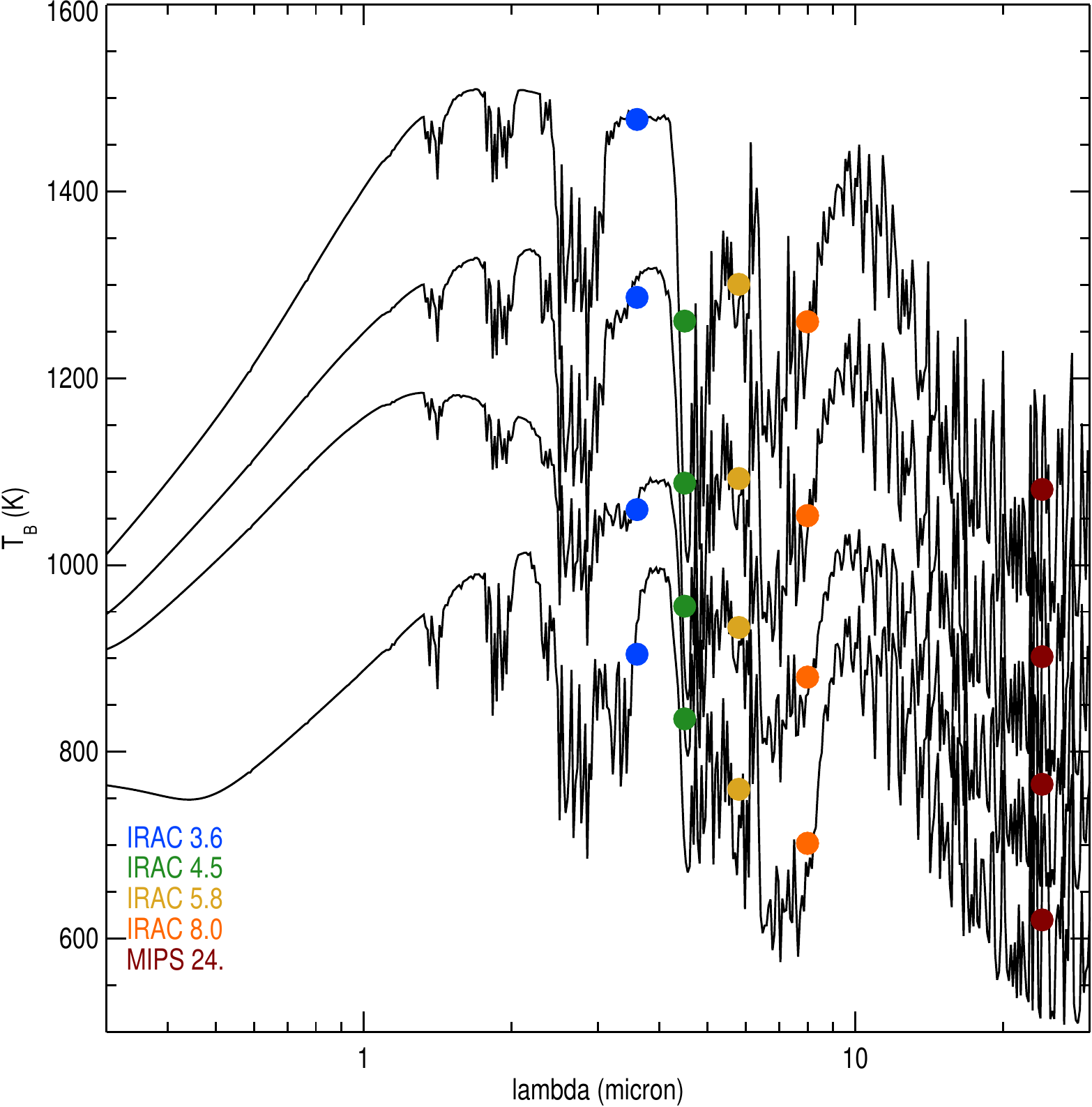}
  \caption{Brightness temperatures as a function of wavelength
    calculated from the results of the radiative hydrodynamics
    simulation for several representative orbital phases. Increasing
    from the bottom, the curves represent observations looking at the
    anti-stellar point, the western terminator, the eastern
    terminator, and the substellar point. Colored dots denote the IRAC
    and MIPS band averaged brightness temperature. The strongly
    non-isothermal nature of the atmosphere at each phase, coupled to
    the strongly wavelength dependent opacity and associated
    contribution function (see text and Figure~\ref{fig:cf}) lead to
    widely varying temperatures.}
  \label{fig:tbright}
\end{figure}

Though we have presented phase amplitudes in terms of flux, one must
be a bit cautious. If the day and night sides of the planet were
isothermal, with hotter and cooler temperatures respectively, the
nature of the blackbody emission would naturally produce a larger
phase amplitude at 3.6~$\mu$m then 4.5~$\mu$m. One potential solution
is to explore the brightness temperature as a function of phase
instead. In Figure~\ref{fig:tbright} we show $T_{bright}$ as a
function of wavelength for several representative phases. The strongly
{\it non}-isothermal nature of the atmosphere in the radial direction
leads to large wavelength dependence in addition to the expected
longitudinal dependence. For this reason, coupled to the fact that
brightness temperature is a more derived quantity leading to larger
uncertainties, we choose to present fluxes. Never the less, as can be
read off the IRAC band-averaged points in the figure, the differential
in brightness temperature between day and night is also larger at
3.6~$\mu$m than at 4.5~$\mu$m. This trend is similarly born out in the
observations of HD~189733b.

\begin{figure*}
  \centering
  \includegraphics[width=0.85\linewidth]{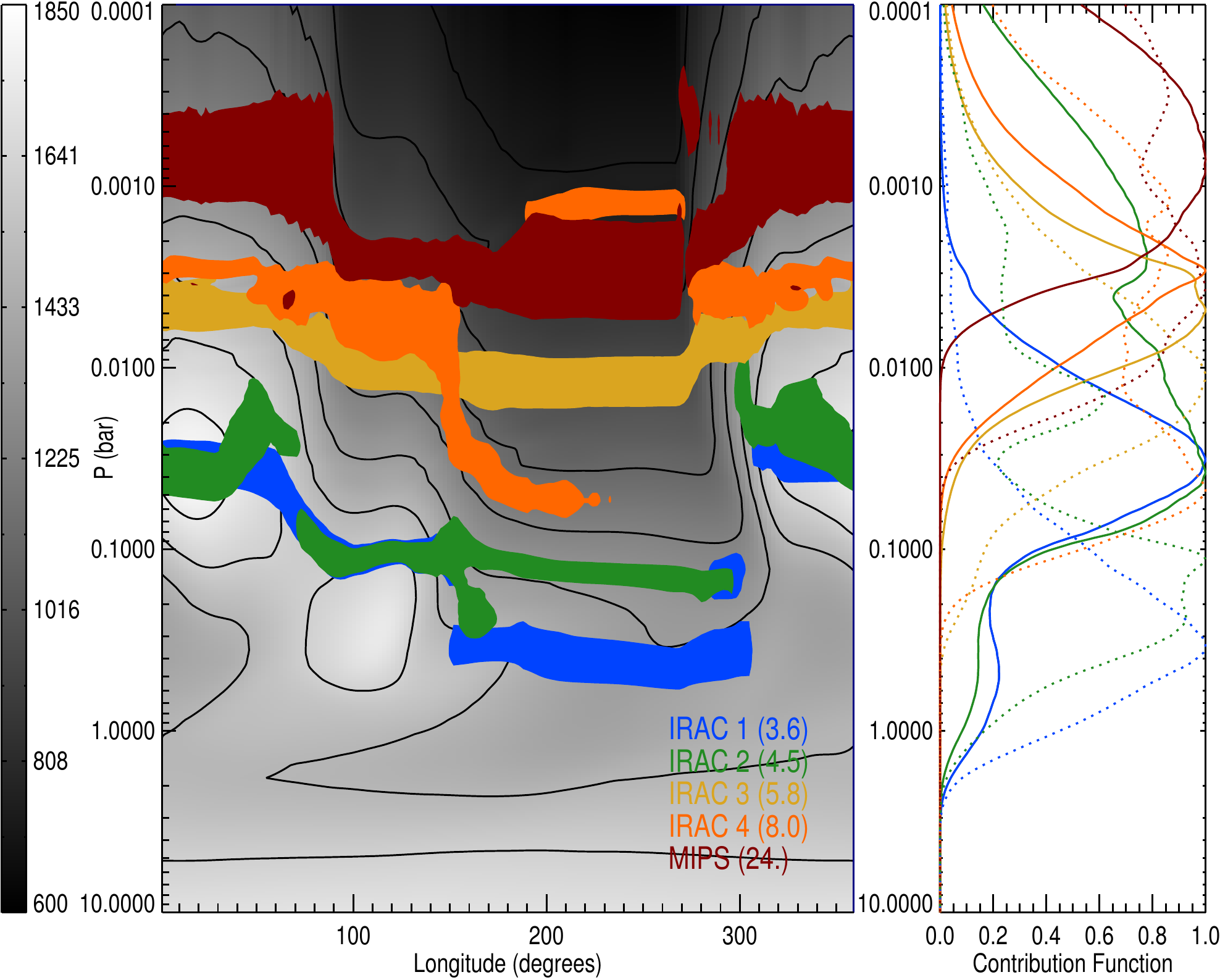}
  \caption{\emph{Left}: The peaks of the normalized contribution
    functions for the IRAC and MIPS bands (colored bands) over-plotted
    on the temperature in K (gray scale) as a function of pressure and
    longitude in degrees along the equator. The sub-stellar longitude
    is at zero. \emph{Right}: The vertical contribution functions at
    the sub-stellar (solid) and anti-stellar (dotted) locations. The
    complex vertical temperature-pressure structure, shaped by the
    underlying radiative hydrodynamics in the model, is clearly a
    strong function of longitude. Assuming thermo-chemical
    equilibrium, this leads to significant variations in the opacities
    of individual bands, in a manner that does not necessary
    correlate.  The result is that the radial location of the
    effective photospheres in the bands do not necessarily maintain
    their relative positions as you move around in longitude.}
  \label{fig:cf}
\end{figure*}

In an attempt to elucidate the mechanism behind the unintuitive
results of Figures~\ref{fig:data}, \ref{fig:theory}, and
\ref{fig:tbright}, we calculate the longitudinally dependent vertical
contribution function at the equator from our model. The contribution
function is given by \citep{chamberlain1987}:
\begin{equation}
cf\left(P,\lambda\right) =
B\left(\lambda,T\right)\frac{d e^{\tau}}{d\log{P}},
\label{eq:cf}
\end{equation}
where $\tau$ is the wavelength dependent optical depth and $B$ is the
local blackbody. The results are shown in Figure~\ref{fig:cf}. On the
left-hand side, the gray scale shows the temperature as a function of
pressure and longitude along the equator, with the sub-stellar point
located at zero degrees longitude. Colored bands indicate the regions
of the atmospheres near the peak of the contribution
functions. The complicated
pressure-longitude structure of the contribution functions is immediately obvious. For
example, the 3.6~$\mu$m photosphere lies at ---or above--- the
4.5~$\mu$m photosphere on the planet's dayside.  On the nightside, on
the other hand, the usual adage ``ch1 probes deeper'' is actually
borne out. 

The convoluted contribution functions arise due to the temperature,
pressure, and wavelength dependence of opacity, which can increase in
one band while simultaneously decreasing in another. The result is
that the effective photospheres in the various bands can cross,
implying that we are probing different respective depths and pressures
in the atmosphere at different orbital phases. The right-hand panel of
Figure~\ref{fig:cf} shows the vertical contribution functions at the
substellar and antistellar points. From these plots it is clear that
the contribution functions are far from delta-functions, but in fact
probe large ranges in pressures and can exhibit several maxima.

There are a number of features in Figure~\ref{fig:cf} that manifest
themselves in Figure~\ref{fig:theory}. Starting with the 5.8, 8.0, and
24~$\mu$m bands, we see that they probe low pressure regions on the
dayside. Given the short radiative timescales at these low pressures,
the hot-spot displacement is quite low resulting in small phase
offsets. The rather low phase offset of the 4.5~$\mu$m point is the
result of two features: the curved nature of the photosphere and the
multiple peaks of the contribution function with height as seen in the
right-hand panel. The curvature of the photosphere means it is probing
cooler temperatures as it moves away from the sub-stellar point,
making the dayside brightness map of the planet more centrally
concentrated (essentially like limb-darkening). The second effect, the
double peaked contribution function, means that a
significant fraction of the emission is coming from higher in the
atmosphere, where again the short radiative timescale tends to reduce
the phase offset. The 3.6~$\mu$m photosphere, on the other hand, is
roughly an isobar on the dayside, hence the greater phase offset at 3.6~$\mu$m.

Phase amplitudes can also be essentially read off
Figure~\ref{fig:cf}. The peak of the contribution function for the
24~$\mu$m band remains high in the atmosphere, where it is cool across
the entire planet, resulting in the smallest phase amplitude. The
phase amplitudes at 5.8 and 8.0~$\mu$m are similar, though the
5.8~$\mu$m crosses more temperature contours, resulting a slightly
larger value. At first glance, the relative phase amplitudes at 4.5
and 3.6~$\mu$m seem to defy the others. However, as can again be seen
in the right-hand panel, the significant contribution at 4.5~$\mu$m
from the upper, cooler regions of the atmosphere produce a lower value
of $F_{\rm max}$ relative to 3.6~$\mu$m and lead to a smaller phase
amplitude.

The convoluted photospheres are due to changes in opacity with
temperature and pressure. The radiative-hydrodynamical models utilize
opacities from \cite{Sharp_Burrows_2007}, which assume solar abundance
and thermo-chemical equilibrium at each temperature and pressure
(i.e.: location) in the atmosphere. The opacities include absorption
due to the four most spectroscopically active species, H$_2$O, CO,
CO$_2$, and CH$_4$. The assumption of thermo-chemical equilibrium is
important, and we discuss below.

\begin{figure*}
  \centering
  \includegraphics[width=0.75\linewidth]{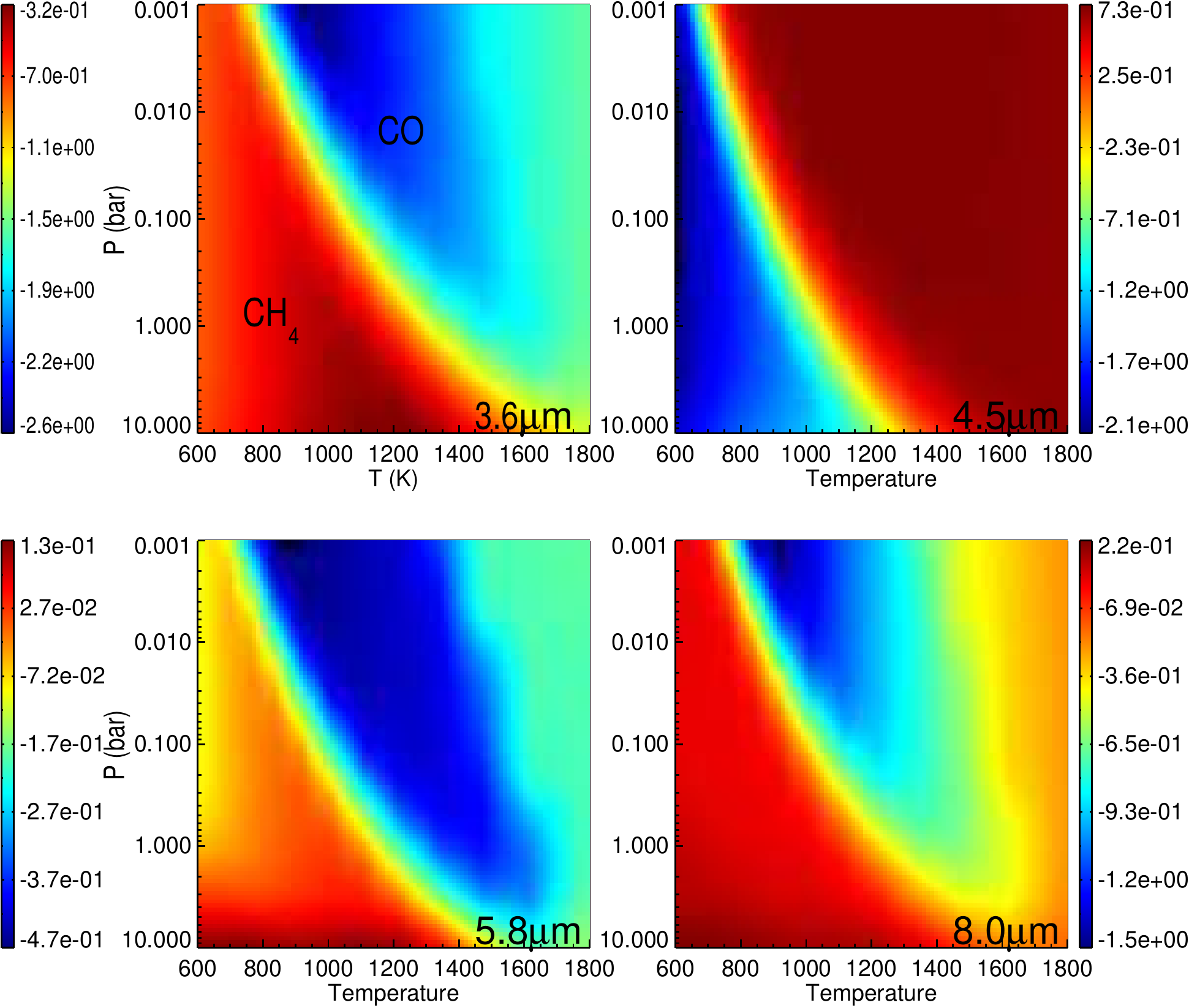}
  \caption{The log of the total opacity (in cm$^2$/g) due to molecules
    as a function of temperature and pressure at 3.6~$\mu$m,
    4.5~$\mu$m, 5.8~$\mu$m, and 8.0~$\mu$m. The opacities are taken
    from \cite{Sharp_Burrows_2007} and assume thermo-chemical
    equilibrium at each temperature and pressure. The obvious feature
    running diagonally through all the plots is the change in opacity
    as the dominate carbon-bearing molecule switches from CH$_4$ in
    the lower left to CO in the upper right. At a given pressure, the
    increase of the opacity at 4.5~$\mu$m with temperature (as opposed
    to the decrease at the other three wavelengths) plays an important
    role in dictating the relative radial location of the photospheres
    with longitude. Note these are not band averaged quantities, but
    instead focus just on the central wavelength of the band.}
  \label{fig:all_opac}
\end{figure*}

To understand the behavior of opacity in the current simulations, we
plot the opacity structure with temperature and pressure at four
representative wavelengths in Figure~\ref{fig:all_opac}. Note that
these are opacities at single wavelengths at the center of the IRAC
bands, not band averaged quantities. The most obvious feature in the
plots is the transition running diagonally in all 4 of the plots,
which corresponds to the CO--CH$_4$ transition. The pressures and
temperatures of HD~189733b happen to straddle this boundary. By
following an isobar across each of the plots one can get a rough idea
of the opacity as a function of longitude near the contribution
function maximum. Of particular note, the opacity at 3.6~$\mu$m
increases by orders of magnitude when moving from day to night, while
at 4.5~$\mu$m it decreases. The fact that they are not correlated
leads directly to the crossing of contribution functions with
longitude ---a given wavelength cannot be assumed to always probe
deeper/shallower than another. Though the carbon chemistry likely
plays an important role in explaining the trend for HD~189733b,
planets with hotter or cooler atmospheres may not straddle this
transition. None the less, as seen in Figure~\ref{fig:data} planets
with a wide range of parameters exhibit this counterintuitive
result. As we discuss below, additional physics, including vertical
mixing and inhomogeneous clouds may result in similar phenomena.

\section{Discussion \& Conclusions}\label{sec:discussion}

\subsection{Clouds \& Disequilibrium Chemistry}
We've used a cloud-free Solar-abundance model where the atmosphere is
everywhere in thermo-chemical equilibrium.  Clouds could in principle
change this story in two ways.  First of all, condensation can reduce
gas phase abundances by orders of magnitude.  But the dominant
absorbers in the Spitzer IRAC channels, H$_2$O, CO, CO$_2$, and
CH$_4$, do not condense anywhere in the atmosphere of a hot Jupiter.
More importantly, mineral condensates increase the opacity, and do so
in a relatively gray fashion.  Uniform clouds would therefore lead to
contribution functions that are less
wavelength-dependent. Inhomogeneous clouds, as have been inferred
based on optical phase curves
\citep{demory2013inference,esteves2015changing,angerhausen2015comprehensive}
and simulations \citep{lee2016}, would complicate the picture, with
cloudy regions having overlapping contribution functions and clear
regions exhibiting the temperature-dependent effects we've discussed
in this paper.

\cite{knutson20123} suggested that the nightside of HD~189733b might
not be in thermo-chemical equilibrium. The observed nightside
brightness temperature is greater at 4.5 than at 3.6~$\mu$m, which
could be explained if CO from deeper layers, or the dayside, were
transported to the vicinity of the IR photosphere faster than the
chemical timescale \citep{Madhusudhan2016}. Indeed, \cite{Cooper2006}
predicted that atmospheric dynamics could mix hot Jupiter atmospheres
faster than chemistry can operate. \cite{Agundez2014} predicted that
HD~189733b should have CO rather than CH$_4$ everywhere in the
atmosphere due to horizontal quenching.

\cite{stevenson2010possible} reported disequilibrium chemistry on
GJ~436b based on its dayside emission spectrum: they saw evidence for
CO rather than CH$_4$.  Since even the substellar photospheric
temperature is too cool for CO, the gas must originate from deeper
layers of the atmosphere, so-called vertical quenching. Vertical
mixing on HD~189733b could also increase the CO abundance at the
nightside photosphere, since at pressures of $\sim$1~bar, CO is
preferred everywhere. Interestingly, at pressures above $\sim$10~bar,
CH$_4$ again becomes the dominant species \citep[e.g., Figure~2
  of][]{Madhusudhan2016}, so vertical mixing cuts both ways.

If hot Jupiter atmospheres are well mixed, then the
temperature-dependent effects we discuss in this Letter are less
dramatic, but still important. In particular, the curious shape of the
4.5~$\mu$m photosphere ---which partially explains the small phase
offsets at this wavelength--- is not driven by carbon chemistry.

In any case, HD~189733b is the coolest and most longitudinally
isothermal hot Jupiter that has been studied to date
\citep{schwartz2015balancing,Schwartz_2017}.  Most hot Jupiters with
published phase curves have much greater longitudinal temperature
contrasts, often measured in thousands, rather than hundreds, of K.
And the very hottest of them have dayside temperatures hot enough to
dissociate molecules and ionize atoms, e.g.\ WASP-12b
\citep{hebb2009wasp}, WASP-33b \citep{Smith2011}, and KELT-9b
\citep{Gaudi2017}.  These transitions should lead to enormous opacity
differences between night and day and hence large excursions in the
wavelength-specific contribution functions. It is not clear to what
extent horizontal and vertical mixing can homogenize the atmospheres
of such worlds.

\subsection{Implications of Non-Isobaric Photospheres}

The wavelength-dependent photospheres are not the concentric spheres
we like to imagine. This does not necessarily impact estimates of
effective temperatures: many schemes to go from brightness
temperatures to an effective temperature are agnostic about the
precise pressures probed by different wavelengths
\citep{cowan2011statistics,schwartz2015balancing}, though energy
budgets based on spectral retrieval may be more sensitive
\citep[e.g.,][]{stevenson2014thermal}. Nor do non-isobaric
photospheres invalidate eclipse and phase mapping, either in a single
broadband, in multiple bands, or even spectral mapping
\citep[limb-darkening could in principle scuttle thermal mapping
  efforts, but it has previously been shown to be
  negligible:][]{cowan2008inverting}.

However, convoluted photospheres do complicate the interpretation of
multi-wavelength maps. In particular, we have shown that automatically
assuming that different wavelengths probe different layers is
incorrect: it is approximately true at any one longitude and latitude
on a planet, but the locations and order of the layers change from one
location to another, primarily driven by differences in temperature
coupled with a temperature-dependent opacity spectrum.

In order to construct multi-dimensional maps of hot Jupiters, we will
have to do something more clever than constructing a layer-cake of
multiple single-wavelength maps. One could instead start by evaluating
the vertical temperature profile at each location on the maps
---especially the regions near the equator that contribute most to the
lightcurves and hence are best-constrained \citep[this step is not the
  same as performing spectral retrieval on disk-integrated spectra as
  was done by][]{stevenson2014thermal}.  One would then combine these
vertical temperature soundings to construct a true 3D map of the
planet's temperature structure \citep[with the usual caveat that the
  latitudinal constraints on the nightside are weak and
  indirect:][]{Majeau_2012,deWit_2012,cowan2013light}.  The challenge
with this approach will be assessing the uncertainty in the brightness
maps at each location in a way that is useful for spectral retrieval
exercises.

Alternatively, it may be possible to fit the disk-integrated
multi-wavelength data simultaneously with a model that accounts not
only for 3D variations in temperature, but also for the wandering
photospheres.  In any case, this is a problem worth tackling soon
given the impending launch of the James Webb Space Telescope.

\acknowledgements
We thank the International Space Science Institute for hosting the
Exo-Cartography workshops.  N.B.C.\ acknowledges support from the
McGill Space Institute, l'Institut de recherche sur les exoplan\`etes,
an NSERC Discovery grant, and a FRQNT Nouveau Chercheur grant. The
authors further thank the anonymous referee for improvements to the
paper.

\bibliographystyle{aasjournal}

\end{document}